\documentclass{elsart}
\usepackage{epsfig}
\usepackage{amssymb}
\begin{document}

\begin{frontmatter}
\title{One-dimensional completed scattering and quantum nonlocality
of entangled states}
\author{N L Chuprikov}

\address{Tomsk State Pedagogical University, 634041, Tomsk, Russia}

\begin{abstract}
Entanglement is usually associated with compound systems. We first show that a
one-dimensional (1D) completed scattering of a particle on a static potential
barrier represents an entanglement of two alternative one-particle sub-processes,
transmission and reflection, macroscopically distinct at the final stage of
scattering. The wave function for the whole ensemble of scattering particles can be
uniquely presented as the sum of two isometrically evolved wave packets to describe
the (to-be-)transmitted and (to-be-)reflected subensembles of particles at all
stages of scattering. A noninvasive Larmor-clock timing procedure adapted to either
subensemble shows that namely the dwell time gives the time spent, on the average,
by a particle in the barrier region, and it denies the Hartman effect. As regards
the group time, it cannot be measured and hence it cannot be accepted as a measure
of the tunneling time. We argue that nonlocality of entangled states appears in
quantum mechanics due to inconsistency of its superposition principle with the
corpuscular properties of a particle. For example, this principle associates a 1D
completed scattering with a single (one-way) process, while a particle, as an
indivisible object, cannot take part in transmission and reflection, simultaneously.
\end{abstract}

\end{frontmatter}

\newcommand{\Api}{A_{in}}
\newcommand{\Ami}{B_{in}}
\newcommand{\Apo}{A_{out}}
\newcommand{\Amo}{B_{out}}
\newcommand{\bpi}{a_{in}}
\newcommand{\bmi}{b_{in}}
\newcommand{\bpo}{a_{out}}
\newcommand{\bmo}{b_{out}}
\newcommand{\api}{a_{in}}
\newcommand{\ami}{b_{in}}
\newcommand{\apo}{a_{out}}
\newcommand{\amo}{b_{out}}

\section{Introduction} \label{aI}

For a long time scattering a particle on one-dimensional (1D) static potential
barriers have been considered in quantum mechanics as a representative of
well-understood phenomena. However, solving the so-called tunneling time problem
(TTP) for a 1D completed scattering (see reviews \cite{Ha2,La1,Olk,Ste,Mu0,Nu0,Ol3}
and references therein) showed that this is not the case.

At present there is a variety of approaches to introduce characteristic times for
the process. They are the group (Wigner) tunneling times (more known as the "phase"
tunneling times) \cite{Ha2,Wig,Har,Ha1,Ter}, different variants of the dwell time
\cite{Ha1,Smi,Ja1,Ja2,But,Le1,Nus,Go1,Mue,Bra}, the Larmor time
\cite{But,Baz,Ryb,Aer,Bu1,Lia,Zhi}, and the concept of the time of arrival which is
based on introducing either a suitable time operator (see, e.g.,
\cite{Aha,Mu4,Hah,Noh}) or the positive operator valued measure \cite{Mu0,Mu9} (see
also \cite{Le3,Le5,Mu5,Le6}). A particular class of approaches to study the temporal
aspects of a 1D scattering includes the Bohmian \cite{Le3,Le2,Gru,Bo1,Kr1}, Feynman
and Wigner ones (see \cite{Sok,Yam,Ymm,Ya1,Kre} as well as \cite{La1,Mu0} and
references therein). One has also point out the papers \cite{Ga1,Ga2,Ga3} to study
the characteristic times of "the forerunner preceding the main tunneling signal of
the wave created by a source with a sharp onset".

The source of a long-lived controversy in solving the TTP, which still persists, is
usually associated with the absence of a Hermitian time operator. However, our
analysis shows that this problem is closely connected to the mystery of quantum
nonlocality of entangled states \cite{Ens,Bel}. As is known, the main peculiarity of
such states is the availability of nonzero correlations between two events separated
with space-like intervals.

The main intrigue is that, though this prediction of quantum theory contradicts
special relativity, now it has been reliably stated (theoretically and
experimentally \cite{Asp}) that nonlocality is indeed an inherent property of
existing quantum mechanics (a deep analysis of this question is done in
\cite{No1,Wis}).

It is now widely accepted that nonlocal correlations of entangled states do not
violate special relativity, for they are not associated with a superluminal
transmission of signals (see, e.g., \cite{Gis,Shi}). However, with regards to this
'no-signalling' interpretation, Bell pointed out that "... we have lost the idea
that correlations can be explained, or at least this idea awaits reformulation. More
importantly, the 'no signaling' notion rests on concepts which are desperately
vague, or vaguely applicable...." (quoted from \cite{No1}).

We agree entirely with this doubt: if nonzero correlations between two events are
not a consequence of a causal relationship between them, then the very notion of
'correlations' becomes physically meaningless. It is just the main challenge of
quantum mechanics that its principles imply introducing such strange correlations.
So that, it is worthwhile to reveal an imperfectness in the foundation of quantum
theory, which creates such a paradoxical situation.

In this paper, the origin of quantum nonlocality is analyzed in the case of a 1D
completed scattering. Studying this particular problem suggests the way of how to
reconcile quantum mechanics with special relativity. We show (Section \ref{a0}) that
existing quantum mechanics does not allow any consistent model of this process. Its
superposition principle, applied to entangled states, contradicts corpuscular
properties of particles. A new, consistent model of a 1D completed scattering, free
of nonlocality, is presented in Sections \ref{a2} and \ref{a3}.

\section{Towards a local model of a 1D completed scattering.}
\label{a0}
\subsection{On the inconsistency of the existing model of a 1D completed
scattering.} \label{a01}

It is evident that a proper theoretical description of any physical phenomenon must
obey the following three requirements which are connected with each other: {\it (i)
it must explain the phenomenon; (ii) it must be consistent; (iii) it can be verified
experimentally}. However, the existing quantum-mechanical model of a 1D completed
scattering does not obey these requirements.

{\bf Firstly}, {\it existing quantum mechanics endows a 1D completed scattering with
quantum nonlocality whose reality is questionable}.

Some manifestations of nonlocality, arose in the existing approaches, have been
pointed out and analyzed by Leavens and co-workers (see \cite{Aer,Le5,Le6,Le2}). For
example, the Bohmian model of a 1D completed scattering predicts that the fate of
the incident particle (to be transmitted or to be reflected by the barrier) depends
on the coordinate of its starting point (see \cite{Le2}). In this case, that of the
critical spatial point to separate the starting regions of to-be-transmitted and
to-be-reflected particles depends on the shape of the potential barrier, though it
is located at a considerable distance from the particle's source.

Further, the time-of-arrival concept \cite{Mu9} predicts a nonzero probability of
arriving a particle at the spatial regions where the probability density is {\it a
priori} zero (see \cite{Le5,Le6}). The Larmor time concept predicts the precession
of the average spin of reflected particles, under the magnetic field localized
beyond the barrier, on the side of transmission where reflected particles are absent
{\it a priori} (see \cite{Aer}).

However, perhaps the most known manifestation of quantum nonlocality, predicted by
the existing model of a 1D completed scattering, is the so-called Hartman effect
(and its versions) which is associated with the anomalously short (or even negative)
times of tunneling a particle through the barrier region (see, e.g.,
\cite{Har,Mu6,Wi1,Ol1,So1,Zh1,So5,Nim,Mar,Wi2,Ran}).

The existing explanations of this effect (see, e.g., \cite{Wi2,Nim}) are made, in
fact, in the spirit of the 'no-signalling' theories. They suggest that anomalously
short dwell and group times do not mean a {\it superluminal} transmission of a
particle through the barrier region. In fact, this means that the notions of the
dwell and group times, as characteristics of the particle's dynamics in the barrier
region, loos theirs initial physical sense.

So, in the existing form, conventional quantum mechanics endows a 1D completed
scattering with quantum nonlocality. Our next step is to show that this prediction
results from inconsistency of the quantum-mechanical principles.

{\bf Secondly}, {\it within the existing framework of quantum mechanics, any
procedure of timing the motion of a scattering particle (both without and with
distinguishing transmission and reflection) is a priory inconsistent}.

On the one hand, the main feature of a particle, as an indivisible object, implies
that it cannot be simultaneously transmitted and reflected by the potential barrier.
So that a 1D scattering should be considered as a combined process to consist from
two alternative sub-processes, transmission and reflection, macroscopically distinct
at the final stage of scattering. And, thus, there should be two experimenters for
studying the subensembles of transmitted or reflected particles.

In this problem, introducing characteristic times and other observables, common for
these two subensembles, has no physical sense. Such quantities simply cannot be
measured, since they describe neither transmitted nor reflected particles. Their
introduction necessitates quantum nonlocality, and they cannot be properly
interpreted (about the interpretation problem for the dwell time, see in
\cite{Ha1,Wi2}). For example, the average value of the particle's position (or
momentum), calculated for the whole ensemble of scattering particles, does not give
the expectation (i.e., most probable) value of this quantity.

On the other hand, the superposition principle, as it stands, demands of treating a
1D completed scattering as a single one-particle process, even at its final stage.
By this principle, the set of one-particle's observables should be introduced namely
for the whole ensemble of scattering particles, i.e., without distinguishing
transmission and reflection.

One has to stress that the existing model of a 1D completed scattering denies, not
only on the conceptual level, introducing individual characteristic times and
observables for transmission and reflection. This model does not provide any
description of these sub-processes at all stages of scattering. All the existing
approaches, which notwithstanding introduce the transmission (or reflection) time,
deal in fact with the subensembles, in which the number of particles is not
conserved.

{\bf Thirdly}, {\it existing quantum mechanics does not allow a consistent procedure
of measuring the time spent by a particle in the barrier region.}

This equally concerns experiments on photonic tunneling which are at present more
reliable than those for electronic tunneling. As is known (see, e.g., \cite{Wi2}),
such experiments imply two steps. At the fist step, a light pulse is sent through a
barrier-free region. The arrival time of the peak of this pulse at a detector is
needed as a reference time. At the second step, an investigated potential barrier is
inserted in the path of the pulse. The arrival time of the transmitted peak at the
detector is measured and then compared with the reference time. The difference of
these two arrival times is considered as a searched-for the group delay time.

The main difficulty of measuring this {\it asymptotic} characteristic time is
usually associated with reshaping the incident light pulse (or, wave packet) in the
barrier region. At the same time there is once more problem which has remained
obscure. It relates to the fact that the above procedure is based on the implicit
assumption that the transmitted and free-evolved peaks start from the same spatial
point.

However, as it follows from our model of a 1D competed scattering, this is not the
case even for the resonant tunneling. So that this procedure gives the time delay
neither for transmitted nor for reflected parts of the incident wave packet. We are
sure that the same is valid for photonic tunneling. Moreover, as will be seen from
our analysis, there is a reason by which the group time is a physical quantity of
secondary importance.

\subsection{How to reconcile a quantum model of a 1D scattering with special
relativity?} \label{a02}

So, as it follows from the above analysis, a principal shortcoming of the existing
quantum model of a 1D completed scattering is that it endows a particle with the
properties to contradict its corpuscular nature.

In this paper we present a new model of this process, which is based on two main
ideas: (i) the state of a particle taking part in a 1D completed scattering is an
entangled (combined) state; (ii) quantum mechanics must distinguish, on the
conceptual level, entangled and unentangled (elementary) states.

We first show that in the problem under consideration, for a given potential and
initial state of a particle, the wave function to describe the whole ensemble of
particles can be uniquely presented as a sum of two isometrically evolved wave
packets which describe alternative sub-processes, transmission and reflection, at
all stages of scattering.

Note, at present all quantum-mechanical rules are equally applied to macroscopically
distinct states and their superpositions. However, the main lesson of solving the
TTP is just that this rule is erroneous. A single system (however, macroscopic or
microscopic) cannot take part simultaneously in two or more macroscopically distinct
sub-processes. This means that the averaging rule (Born's formula) is not applicable
to entangled states.

The phenomenon of quantum nonlocality results from ignoring this restriction. In
other words, it appears when one attempts to associate, contrary to the nature of
entangled states, the interference pattern formed by a superposition of
macroscopically distinct sub-processes with a single causally evolved process.

\newcommand{\ko}{\kappa_0^2}
\newcommand{\kj}{\kappa_j^2}
\newcommand{\kd}{\kappa_j d_j}
\newcommand{\kki}{\kappa_0\kappa_j}

\newcommand{\Ra}{R_{j+1}}
\newcommand{\Rb}{R_{(1,j)}}
\newcommand{\Rc}{R_{(1,j+1)}}

\newcommand{\Ta}{T_{j+1}}
\newcommand{\Tb}{T_{(1,j)}}
\newcommand{\Tc}{T_{(1,j+1)}}

\newcommand{\Wa}{w_{j+1}}
\newcommand{\Wb}{w_{(1,j)}}
\newcommand{\Wc}{w_{(1,j+1)}}

\newcommand{\UU}{u^{(+)}_{(1,j)}}
\newcommand{\VV}{u^{(-)}_{(1,j)}}

\newcommand{\ta}{t_{j+1}}
\newcommand{\tb}{t_{(1,j)}}
\newcommand{\tc}{t_{(1,j+1)}}

\newcommand{\tee}{\vartheta_{(1,j)}}

\newcommand{\tta}{\tau_{j+1}}
\newcommand{\ttb}{\tau_{(1,j)}}
\newcommand{\ttc}{\tau_{(1,j+1)}}

\newcommand{\FF}{\chi_{(1,j)}}
\newcommand {\aro}{(k)}
\newcommand {\da}{\partial}
\newcommand{\ppp}{\mbox{\hspace{5mm}}}
\newcommand{\ooo}{\mbox{\hspace{3mm}}}
\newcommand{\ooa}{\mbox{\hspace{1mm}}}

\section{Wave functions for transmission and reflection}\label{a2}
\subsection{Setting the problem} \label{a1}

Let us consider a particle incident from the left on the static potential barrier
$V(x)$ confined to the finite spatial interval $[a,b]$ $(a>0)$; $d=b-a$ is the
barrier width. Let its in-state, $\psi_{in}(x),$ at $t=0$ be a normalized function
to belong to the set $S_{\infty}$ consisting from infinitely differentiable
functions vanishing exponentially in the limit $|x|\to \infty$. The
Fourier-transform of such functions are known to belong to the set $S_{\infty},$
too. In this case the position, $\hat{x},$ and momentum, $\hat{p},$ operators both
are well-defined. Without loss of generality we will suppose that
\begin{eqnarray} \label{444}
<\psi_{in}|\hat{x}|\psi_{in}>=0,\ooo <\psi_{in}|\hat{p}|\psi_{in}> =\hbar k_0 >
0,\ooo <\psi_{in}|\hat{x}^2|\psi_{in}> =l_0^2;
\end{eqnarray}
here $l_0$ is the wave-packet's half-width at $t=0$ ($l_0<<a$).

We consider a completed scattering. This means that the average velocity, $\hbar
k_0/m,$ is large enough, so that the transmitted and reflected wave packets do not
overlap each other at late times. As for the rest, the relation of the average
energy of a particle to the barrier's height may be any by value.

We begin our analysis with the derivation of expressions for the incident,
transmitted and reflected wave packets to describe, in the problem at hand, the
whole ensemble of particles. For this purpose we will use the variant (see
\cite{Ch1}) of the well-known transfer matrix method \cite{Mez}. Let the wave
function $\psi_{full}(x,k)$ to describe the stationary state of a particle in the
out-of-barrier regions be written in the form
\begin{eqnarray} \label{1}
\psi_{full}(x;k)=e^{ikx}+b_{out}(k)e^{ik(2a-x)}, \ooo for \ooo x\le a;
\end{eqnarray}
\begin{eqnarray} \label{2}
\psi_{full}(x;k)=a_{out}(k)e^{ik(x-d)}, \ooo for \ooo x>b.
\end{eqnarray}
\newcommand{\iii}{\mbox{\hspace{10mm}}}
The coefficients entering this solution are connected by the transfer matrix ${\bf
Y}$:
\begin{eqnarray} \label{50}
\left(\begin{array}{c} 1 \\ b_{out}e^{2ika}
\end{array} \right)={\bf Y} \left(\begin{array}{c} a_{out}e^{-ikd} \\ 0
\end{array} \right), \ppp
{\bf Y}=\left(\begin{array}{cc} q & p \\ p^* & q^* \end{array} \right);
\end{eqnarray}
\begin{eqnarray} \label{500}
q=\frac{1}{\sqrt{T(k)}}\exp\left[i(kd-J(k))\right],\ooa
p=\sqrt{\frac{R(k)}{T(k)}}\exp\left[i\left(\frac{\pi}{2}+ F(k)-ks\right)\right]
\end{eqnarray}
\noindent where $T$, $J$ and $F$ are the real tunneling parameters: $T(k)$ (the
transmission coefficient) and $J(k)$ (phase) are even and odd functions of $k$,
respectively; $F(-k)=\pi-F(k)$; $R(k)=1-T(k)$; $s=a+b$. We will suppose that the
tunneling parameters have already been calculated.

In the case of many-barrier structures, for this purpose one may use the recurrence
relations obtained in \cite{Ch1} just for these real parameters. For the rectangular
barrier of height $V_0$,
\begin{eqnarray} \label{501}
T=\left[1+\vartheta^2_{(+)}\sinh^2(\kappa d)\right]^{-1},\ooo
J=\arctan\left(\vartheta_{(-)}\tanh(\kappa d)\right),\\F=0,\ooo
\kappa=\sqrt{2m(V_0-E)}/\hbar,\nonumber
\end{eqnarray}
if $E<V_0$; and
\begin{eqnarray} \label{502}
T=\left[1+\vartheta^2_{(-)}\sin^2(\kappa d)\right]^{-1},\ooo
J=\arctan\left(\vartheta_{(+)}\tan(\kappa d)\right),\\
F=\left\{\begin{array}{c} 0,\ooo if \ooo \vartheta_{(-)}\sin(\kappa d)\geq 0 \\
\pi,\ooo otherwise,
\end{array} \right.\ooo
\kappa=\sqrt{2m(E-V_0)}/\hbar,\nonumber
\end{eqnarray}
if $E\geq V_0$; in both cases
$\vartheta_{(\pm)}=\frac{1}{2}\left(\frac{k}{\kappa}\pm \frac{\kappa}{k}\right)$
(see \cite{Ch1}).

Now, taking into account Exps. (\ref{50}) and (\ref{500}), we can write down
in-asymptote $\psi_{in}(x,t)$ and out-asymptote $\psi_{out}(x,t)$ for the
time-dependent scattering problem (see \cite{Tei}):
\begin{eqnarray} \label{59}
\psi_{in}(x,t)=\frac{1}{\sqrt{2\pi}}\int_{-\infty}^{\infty} f_{in}(k,t)
e^{ikx}dk,\ooo f_{in}=\Api(k)\exp[-i E(k)t/\hbar];
\end{eqnarray}
\begin{eqnarray} \label{60}
\psi_{out}(x,t)=\frac{1}{\sqrt{2\pi}}\int_{-\infty}^{\infty} f_{out}(k,t)
e^{ikx}dk,\ooo f_{out}= f_{out}^{tr}+f_{out}^{ref};
\end{eqnarray}
\begin{eqnarray} \label{61}
f_{out}^{tr}=\sqrt{T(k)}\Api(k) \exp[i(J(k)-kd-E(k)t/\hbar)]
\end{eqnarray}
\begin{eqnarray} \label{62}
f_{out}^{ref}=\sqrt{R(k)}\Api(-k)\exp[-i(J(k)-F(k)-\frac{\pi}{2}+2ka+E(k)t/\hbar)];
\end{eqnarray}
where Exps. (\ref{59}), (\ref{61}) and (\ref{62}) describe, respectively, the
incident, transmitted and reflected wave packets. Here $\Api(k)$ is the
Fourier-transform of $\psi_{in}(x).$ For example, for the Gaussian wave packet to
obey condition (\ref{444}), $\Api(k)=c\cdot \exp(-l_0^2(k-k_0)^2);$ $c$ is a
normalization constant.

Let us now show that by the final states (\ref{60})-(\ref{62}) one can uniquely
reconstruct the prehistory of the subensembles of transmitted and reflected
particles, impinging the barrier from the left, at all stages of scattering. Let
$\psi_{tr}(x,t)$ and $\psi_{ref}(x,t)$ be searched-for wave packets to describe
transmission (TWF) and reflection (RWF), respectively. By our approach their sum
should give the (full) wave function $\psi_{full}(x,t)$ to describe the whole 1D
completed scattering:
\begin{eqnarray} \label{261}
\psi_{full}(x,t)=\psi_{tr}(x,t)+\psi_{ref}(x,t).
\end{eqnarray}

\subsection{Incoming waves for transmission and reflection}\label{a21}

We begin our analysis with searching for the stationary wave functions for
reflection, $\psi_{ref}(x;k),$ and transmission, $\psi_{tr}(x;k),$ in the region $x
\le a$. Let us write down these two solutions to the stationary Schr\"odinger
equation in the form,
\begin{eqnarray} \label{265}
\psi_{ref}(x;k)=\Api^{ref}e^{ikx}+\Amo^{ref}e^{-ikx},\ooo \psi_{tr}(x;k)=
\Api^{tr}e^{ikx}+\Amo^{tr}e^{-ikx},
\end{eqnarray}
where coefficients obey the following conditions:
\begin{eqnarray} \label{2650}
\Api^{tr}+\Api^{ref}=1;\ooo \Amo^{tr}=0;\ooo \Amo^{ref}=b_{out}e^{2ika}.
\end{eqnarray}
Besides, we suppose that reflected particles do not cross the barrier region, and,
hence, the probability flux for $\psi_{ref}(x;k)$ should be equal to zero:
\begin{eqnarray} \label{264}
|\Api^{ref}|^2=|b_{out}|^2\equiv R(k).
\end{eqnarray}
By the same reason, the probability flux for $\psi_{full}(x;k)$ and $\psi_{tr}(x;k)$
should be the same,
\begin{eqnarray}\label{263}
|\Api^{tr}|^2=|a_{out}|^2\equiv T(k)
\end{eqnarray}
Taking into account that $\psi_{tr}=\psi_{full}-\psi_{ref},$ we can exclude
$\psi_{tr}$ from Eq. (\ref{263}). As a result, we obtain
\begin{eqnarray} \label{2630}
\Re\left(\Api^{ref} \right)-|\amo|^2=0.
\end{eqnarray}
Thus, from Eqs. (\ref{264}) and (\ref{2630}) it follows that
$\Api^{tr}=\sqrt{T}(\sqrt{T}\mp i\sqrt{R})$; $\Api^{ref}=\sqrt{R}(\sqrt{R}\pm
i\sqrt{T}) \equiv \sqrt{R}\exp(i\lambda)$; $\lambda=\pm\arctan(\sqrt{T/R})$.

So, a coherent superposition of the incoming waves to describe transmission and
reflection, for a given $E$, yields the incoming wave of unite amplitude, that
describes the whole ensemble of incident particles. In this case, not only
$\Api^{tr}+\Api^{ref}=1$, but also $|\Api^{tr}|^2+|\Api^{ref}|^2=1$! Besides, the
phase difference for the incoming waves to describe reflection and transmission
equals $\pi/2$ irrespective of the value of $E$.

Our next step is to show that only one root of $\lambda$ gives a searched-for
$\psi_{ref}(x;k).$ For this purpose the above solution should be extended into the
region $x>a$. To do this, we will further restrict ourselves by symmetric potential
barriers, though the above analysis is valid in the general case.

\subsection{Wave functions for transmission and reflection in the region of a
symmetric potential barrier}\label{a22}

Let $V(x)$ be such that $V(x-x_c)=V(x_c-x);$ $x_c=(a+b)/2.$ As is known, for the
region of a symmetric potential barrier, one can always find odd, $u(x-x_c)$, and
even, $v(x-x_c)$, solutions to the Schr\"odinger equation. We will suppose here that
these functions are known. For example, for the rectangular potential barrier (see
Exps. (\ref{501}) and (\ref{502})),
\[ u(x)=\sinh(\kappa x),\ooo v(x)=\cosh(\kappa x),\ooo if\ooo E\le V_0;\]
\[ u(x)=\sin(\kappa x),\ooo v(x)=\cos(\kappa x),\ooo if\ooo E\ge V_0.\]
Note, $\frac{du}{dx}v-\frac{dv}{dx}u$ is a constant, which equals $\kappa$ in the
case of the rectangular barrier. Without loss of generality we will keep this
notation for any symmetric potential barrier.

Before finding $\psi_{ref}(x;k)$ and $\psi_{tr}(x;k)$ in the barrier region, we have
firstly to derive expressions for the tunneling parameters of symmetric barriers.
Let in the barrier region $\psi_{full}(x;k)=a_{full}\cdot u(x-x_c,k)+b_{full}\cdot
v(x-x_c,k).$ "Sewing" this expression together with Exps. (\ref{1}) and (\ref{2}) at
the points $x=a$ and $x=b$, respectively, we obtain
\begin{eqnarray*}
a_{full}=\frac{1}{\kappa}\left(P+P^*b_{out}\right)e^{ika}=
-\frac{1}{\kappa}P^*a_{out}e^{ika};\nonumber\\
b_{full}=\frac{1}{\kappa}\left(Q+Q^*b_{out}\right)e^{ika}=
\frac{1}{\kappa}Q^*a_{out}e^{ika};\nonumber
\end{eqnarray*}
\begin{eqnarray*}
Q=\left(\frac{du(x-x_c)}{dx}+i k u(x-x_c)\right)\Bigg|_{x=b};\nonumber\\
P=\left(\frac{dv(x-x_c)}{dx}+i k v(x-x_c)\right)\Bigg|_{x=b}.\nonumber
\end{eqnarray*}
As a result,
\begin{eqnarray} \label{300}
a_{out}=\frac{1}{2}\left(\frac{Q}{Q^*}-\frac{P}{P^*}\right);\ooo
b_{out}=-\frac{1}{2}\left(\frac{Q}{Q^*}+\frac{P}{P^*}\right).
\end{eqnarray}
As it follows from (\ref{50}), $a_{out}=\sqrt{T}\exp(iJ),$
$b_{out}=\sqrt{R}\exp\left(i\left(J-F-\frac{\pi}{2}\right)\right)$. Hence
$T=|a_{out}|^2,$ $R=|b_{out}|^2,$ $J=\arg(a_{out})$. Besides, for symmetric
potential barriers $F=0$ when $\Re(QP^*)>0$; otherwise, $F=\pi$.

Then, one can show that "sewing" the general solution $\psi_{ref}(x;k)$ in the
barrier region together with Exp. (\ref{265}) at $x=a$, for both the roots of
$\lambda$, gives odd and even functions in this region. For the problem considered,
only the former has a physical meaning. The corresponding roots for $\Api^{ref}$ and
$\Api^{tr}$ read as
\begin{eqnarray} \label{301}
\Api^{ref}=b_{out}\left(b^*_{out}-a^*_{out}\right);\ooa
\Api^{tr}=a^*_{out}\left(a_{out}+b_{out}\right)
\end{eqnarray}
One can easily show that in this case
\begin{eqnarray} \label{302}
\frac{Q^*}{Q}=-\frac{\Api^{ref}}{b_{out}}=\frac{\Api^{tr}}{a_{out}};
\end{eqnarray}
for $a\le x\le b$
\begin{eqnarray} \label{3000}
\psi_{ref}=\frac{1}{\kappa}\left(PA_{in}^{ref}+P^*b_{out}\right)e^{ika}u(x-x_c,k).
\end{eqnarray}
The extension of this solution onto the region $x\ge b$ gives
\begin{eqnarray} \label{3001}
\psi_{ref}=-b_{out}e^{ik(x-d)}-\Api^{ref}e^{-ik(x-s)}.
\end{eqnarray}

So, Exps. (\ref{265}), (\ref{3000}) and (\ref{3001}) give the solution to the
Schr\"odinger equation, which we expect to describe reflection. Then, the
corresponding solution for transmission is
$\psi_{tr}(x;k)=\psi_{full}(x;k)-\psi_{ref}(x;k).$

Note that $\psi_{full}(x;k)$ does not contain an incoming wave impinging the barrier
from the right, while the found TWF and RWF include such waves. That is, the
superposition of these probability waves leads, due to interference, to their
macroscopical reconstruction: in the superposition, both outgoing waves are
connected only with the left source of particles. One can show that, in this case,
the reflected and transmitted waves are connected causally with the incoming waves
$\Api^{ref}e^{ikx}$ and $\Api^{tr}e^{ikx}$, respectively.

Indeed, let us firstly consider reflection. As is seen from Exp. (\ref{3000}),
$\psi_{ref}(x;k)$ is equal to zero at the point $x_c$, for all values of $k$. As a
result, the probability flux, for any time-dependent wave function formed from
$\psi_{ref}(x;k)$, is equal to zero at this point, for any value of time. This means
that, in the case of reflection, particles impinging the symmetric barrier from the
left do not enter the region $x\ge x_c$. In other words, the wave packet,
$\tilde{\psi}_{ref}(x;k),$ to describe such particles can be written in the form
\begin{eqnarray} \label{3002}
\tilde{\psi}_{ref}(x;k)\equiv \psi_{ref}(x;k)\ooo for\ooo  x\le x_c; \ppp
\tilde{\psi}_{ref}(x;k)\equiv 0 \ooo for\ooo x\ge x_c.
\end{eqnarray}

Note, for a given potential, $\tilde{\psi}_{ref}(x;k)$ does not obey the
Schr\"odinger equation at the point $x=x_c$. Nevertheless the probability density
for this function is everywhere continuous and the probability flux is everywhere
equal to zero. This means that the wave packet, $\tilde{\psi}_{ref}(x,t),$ formed
from the functions $\tilde{\psi}_{ref}(x;k)$ with different $k$, despite
discontinuity its first derivation at the point $x_c$, is everywhere continuous and
evolves with a fixed norm. As is said above, namely this packet describes the
subensemble of particles which impinge the barrier from the left and are reflected
by it.

The above suggests that the subensemble of incident particles to be transmitted by
the barrier is described by the incident wave $\Api^{tr}e^{ikx}$ of the solution
$\psi_{tr}(x;k).$ Namely this incident wave is causally connected with the
transmitted one $a_{out}(k)e^{ik(x-d)}$ of the solution $\psi_{full}(x;k)$.

One can easily show that the function $\tilde{\psi}_{tr}(x;k),$ where
$\tilde{\psi}_{tr}(x;k)=\psi_{full}(x;k) -\tilde{\psi}_{ref}(x;k),$ is everywhere
continuous and the corresponding probability flux is everywhere constant. In this
case,
\begin{eqnarray} \label{3003}
\tilde{\psi}_{tr}(x;k)\equiv \psi_{tr}(x;k)\ooa for\ooa  x\le x_c; \ooo
\tilde{\psi}_{tr}(x;k)\equiv \psi_{full}(x;k) \ooa for\ooa x\ge x_c.
\end{eqnarray}

As in the case of reflection, the wave packet, $\tilde{\psi}_{tr}(x,t),$ formed from
the functions $\tilde{\psi}_{tr}(x;k)$ (despite discontinuity its first derivation
at the point $x_c$) is everywhere continuous and evolves with a fixed norm. Hence
namely this packet describes the subensemble of particles which impinge the barrier
from the left and are transmitted by it.

One can easily show that
\begin{eqnarray} \label{303}
\tilde{\psi}_{tr}(x;k)=a^l_{tr}u(x-x_c,k)+b_{tr}v(x-x_c,k)\ooo for\ooo a\le x\le
x_c;
\end{eqnarray}
\begin{eqnarray} \label{304}
\tilde{\psi}_{tr}(x;k)=a^r_{tr}u(x-x_c,k)+b_{tr}v(x-x_c,k)\ooo for\ooo x_c\le x\le
b;
\end{eqnarray}
\begin{eqnarray} \label{305}
\tilde{\psi}_{tr}(x;k)=a_{out}e^{ik(x-d)}\ooo for\ooo x\ge b.;
\end{eqnarray}
where
\begin{eqnarray*}
a^l_{tr}=\frac{1}{\kappa}PA_{in}^{tr}e^{ika},\ooa
b_{tr}=b_{full}=\frac{1}{\kappa}Q^*a_{out}e^{ika},\ooa
a^r_{tr}=a_{full}=-\frac{1}{\kappa}P^*a_{out}e^{ika}
\end{eqnarray*}

Note, for any value of $t$
\begin{eqnarray*}  {\bf T}=<\tilde{\psi}_{tr}(x,t)|\tilde{\psi}_{tr}(x,t)>=const;\ooa {\bf
R}=<\tilde{\psi}_{ref}(x,t)|\tilde{\psi}_{ref}(x,t)>=const;
\end{eqnarray*}
${\bf T}$ and ${\bf R}$ are the average transmission and reflection coefficients,
respectively. Besides,
\begin{eqnarray} \label{700100}
<\psi_{full}(x,t)|\psi_{full}(x,t)> ={\bf T}+{\bf R}=1.
\end{eqnarray}
From this it follows, in particular, that the scalar product of the wave packets for
transmission and reflection, $<\tilde{\psi}_{tr}(x,t)|\tilde{\psi}_{ref}(x,t)>,$ is
a purely imagine quantity to approach zero when $t\to\infty$.

We have to stress that these wave packets are not solutions to the Schr\"odinger
equation for a given potential, just as transmission and reflection described by
them are not independent quantum processes. These wave packets may be considered
only as parts of an entangled state to describe a 1D completed scattering, like the
sub-processes may be considered only as two different alternatives to constitute the
same one-particle scattering process.

Of importance is that namely these two wave packets describe the (to-be-)transmitted
and (to-be-)reflected subensembles of particles at all stages of scattering. In this
case
\begin{eqnarray} \label{2610}
\psi_{full}(x,t)=\psi_{tr}(x,t)+\psi_{ref}(x,t)=
\tilde{\psi}_{ref}(x,t)+\tilde{\psi}_{ref}(x,t).
\end{eqnarray}
(Below we will deal only with $\tilde{\psi}_{ref}$ and $\tilde{\psi}_{tr}.$ For this
reason these notations will be used without tilde.)

Now we can proceed to the study of temporal aspects of a 1D completed scattering.
The found wave packets for transmission and reflection permit us to introduce
characteristic times for either sub-process. As will be seen from the following, the
motion of either subensemble of particles in the barrier region can be investigated
with help of the Larmor-clock timing procedure adapted to the sub-processes.

\section{Characteristic times for transmission and reflection} \label{a3}

So, our main purpose now is to find, for each sub-process, the time spent, on the
average, by a particle in the barrier region. In doing so, we have to take into
account that a chosen timing procedure must not influence an original value of the
characteristic time.

Under such conditions, perhaps, the only way to measure the tunneling time for a
completed scattering is to exploit internal degrees of freedom of quantum particles.
As is known, namely this idea underlies the Larmor-time concept based on the Larmor
precession of the particle's spin under the infinitesimal magnetic field.

However, as will be seen from the following, the Larmor-time concept is directly
connected to the dwell time to describe the stationary scattering problem. By this
reason, we define firstly the dwell times for transmission and reflection for a
particle in the stationary state.

\subsection{Dwell times for transmission and reflection}\label{a32}

Note, in the case of transmission the density of the probability flux, $I_{tr}$, for
$\psi_{tr}(x;k)$ is everywhere constant and equal to $T\cdot\hbar k/m$. The
velocity, $v_{tr}(x,k)$, of an infinitesimal element of the flux, at the point $x,$
equals $v_{tr}(x)=I_{tr}/|\psi_{tr}(x;k)|^2.$ Outside the barrier region the
velocity is everywhere constant: $v_{tr}=\hbar k/m$. In the barrier region it
depends on $x$. In the case of an opaque rectangular potential barrier, $v_{tr}(x)$
decreases exponentially when the infinitesimal element approaches the midpoint
$x_c$. One can easily show that $|\psi_{tr}(a;k)|=|\psi_{tr}(b;k)|=\sqrt{T}$, but
$|\psi_{tr}(x_c;k)|\sim\sqrt{T}\exp(\kappa d/2)$.

Thus, any selected infinitesimal element of the flux passes the barrier region for
the time $\tau^{tr}_{dwell}$, where
\begin{eqnarray} \label{4005}
\tau^{tr}_{dwell}(k)=\frac{1}{I_{tr}}\int_a^b|\psi_{tr}(x;k)|^2 dx.
\end{eqnarray}
By analogy with \cite{But} we will call this time scale the dwell time for
transmission.

For the rectangular barrier this time reads (for $E< V_0$ and $E\ge V_0$,
respectively) as
\begin{eqnarray} \label{4007}
\tau^{tr}_{dwell}=\frac{m}{2\hbar k\kappa^3}\left[\left(\kappa^2-k^2\right)\kappa d
+\kappa_0^2 \sinh(\kappa d)\right],
\end{eqnarray}
\begin{eqnarray} \label{4009}
\tau^{tr}_{dwell}=\frac{m}{2\hbar k\kappa^3}\left[\left(\kappa^2+k^2\right)\kappa d
-\beta \kappa_0^2 \sin(\kappa d)\right].
\end{eqnarray}

In the case of reflection the situation is less simple. The above arguments are not
applicable here, for the probability flux for $\psi_{ref}(x,k)$ is zero. However, as
is seen, the dwell time for transmission coincides, in fact, with Buttiker's dwell
time introduced however on the basis of the wave function for transmission.
Therefore, making use of the arguments by Buttiker, let us define the dwell time for
reflection, $\tau^{ref}_{dwell}$, as
\begin{eqnarray} \label{40014}
\tau^{ref}_{dwell}(k)=\frac{1}{I_{ref}} \int_a^{x_c}|\psi_{ref}(x,k)|^2 dx;
\end{eqnarray}
where $I_{ref}=R\cdot \hbar k/m$ is the incident probability flux for reflection.

Again, for the rectangular barrier
\begin{eqnarray} \label{40030}
\tau^{ref}_{dwell}=\frac{m k}{\hbar \kappa}\cdot\frac{\sinh(\kappa d)-\kappa
d}{\kappa^2+\kappa^2_0 \sinh^2(\kappa d/2)}\ooo for \ooo E<V_0;
\end{eqnarray}
\begin{eqnarray} \label{40031}
\tau^{ref}_{dwell}=\frac{m k}{\hbar \kappa}\cdot\frac{\kappa d-\sin(\kappa
d)}{\kappa^2+\beta\kappa^2_0 \sin^2(\kappa d/2)}\ooo for \ooo E\ge V_0.
\end{eqnarray}
As is seen, for rectangular barriers the dwell times for transmission and reflection
do not coincide with each other, unlike the asymptotic group times.

We have to stress once more that Exps. (\ref{4005}) and (\ref{40014}), unlike
Smith's, Buttiker's and Bohmian dwell times, are defined in terms of the TWF and
RWF. As will be seen from the following, the dwell times introduced can be justified
in the framework of the Larmor-time concept.

\subsection{Larmor times for transmission and reflection}\label{a33}

As was said above, both the group (see \cite{Ch5}) and dwell time concepts do not
give the way of measuring the time spent by a particle in the barrier region. This
task can be solved in the framework of the Larmor time concept. As is known, the
idea to use the Larmor precession as clocks was proposed by Baz' \cite{Baz} and
developed later by Rybachenko \cite{Ryb} and B\"{u}ttiker \cite{But} (see also
\cite{Aer,Lia}).

However, we have to stress that the existing concept of the Larmor time was
introduced on the basis of incoming and outgoing waves (see \cite{But,Aer,Lia}). In
this connection, our next step is to redefine the Larmor times for the barrier
region, making use the expressions of the corresponding wave functions just for this
region.

\subsubsection{Preliminaries} \label{a330}

Let us consider the quantum ensemble of electrons moving along the $x$-axis and
interacting with the symmetrical time-independent potential barrier $V(x)$ and small
magnetic field (parallel to the $z$-axis) confined to the finite spatial interval
$[a,b].$ Let this ensemble be a mixture of two parts. One of them consists from
electrons with spin parallel to the magnetic field. Another is formed from particles
with antiparallel spin.

Let at $t=0$ the in state of this mixture be described by the spinor
\begin{eqnarray} \label{9001}
\Psi_{in}(x)=\frac{1}{\sqrt{2}}\left(\begin{array}{c} 1 \\ 1
\end{array} \right)\psi_{in}(x),
\end{eqnarray}
where $\psi_{in}(x)$ is a normalized function to satisfy conditions (\ref{444}). So
that we will consider the case, when the spin coherent in state (\ref{9001}) is the
eigenvector of $\sigma_x$ with the eigenvalue 1 (the average spin of the ensemble of
incident particles is oriented along the $x$-direction); hereinafter, $\sigma_x,$
$\sigma_y$ and $\sigma_z$ are the Pauli spin matrices.

For electrons with spin up (down), the potential barrier effectively decreases
(increases), in height, by the value $\hbar\omega_L/2$; here $\omega_L$ is the
frequency of the Larmor precession; $\omega_L=2\mu B/\hbar,$ $\mu$ denotes the
magnetic moment. The corresponding Hamiltonian has the following form,
\begin{eqnarray} \label{900200}
\hat{H}=\frac{\hat{p}^2}{2m}+V(x)-\frac{\hbar\omega_L}{2}\sigma_z, \ooo if\ooo
x\in[a,b];\ooa \hat{H}=\frac{\hat{p}^2}{2m}, \ooo otherwise.
\end{eqnarray}
For $t>0$, due to the influence of the magnetic field, the states of particles with
spin up and down become different. The probability to pass the barrier is different
for them. Let for any value of $t$ the spinor to describe the state of particles
read as
\begin{eqnarray} \label{9002}
\Psi_{full}(x,t)=\frac{1}{\sqrt{2}}\left(\begin{array}{c}
\psi_{full}^{(\uparrow)}(x,t) \\
\psi_{full}^{(\downarrow)}(x,t) \end{array} \right).
\end{eqnarray}

In accordance with (\ref{261}) (or (\ref{2610})), either spinor component can be
uniquely presented as a coherent superposition of two probability fields to describe
transmission and reflection:
\begin{eqnarray} \label{9003}
\psi_{full}^{(\uparrow\downarrow))}(x,t)=
\psi_{tr}^{(\uparrow\downarrow))}(x,t)+\psi_{ref}^{(\uparrow\downarrow))}(x,t);
\end{eqnarray}
note that $\psi_{ref}^{(\uparrow\downarrow)}(x,t)\equiv 0$ for $x\ge x_c$. As a
consequence, the same decomposition takes place for spinor (\ref{9002}):
$\Psi_{full}(x,t)= \Psi_{tr}(x,t)+\Psi_{ref}(x,t).$

We will suppose that all the wave functions for transmission and reflection are
known. It is important to stress here (see (\ref{700100}) that
\begin{eqnarray} \label{900100}
<\psi_{full}^{(\uparrow\downarrow)}(x,t)|\psi_{full}^{(\uparrow\downarrow)}(x,t)>
=T^{(\uparrow\downarrow)}+R^{(\uparrow\downarrow)}=1;\\
T^{(\uparrow\downarrow)}=<\psi_{tr}^{(\uparrow\downarrow)}(x,t)|
\psi_{tr}^{(\uparrow\downarrow)}(x,t)>=const;\nonumber\\
R^{(\uparrow\downarrow)}=<\psi_{ref}^{(\uparrow\downarrow)}(x,t)|
\psi_{ref}^{(\uparrow\downarrow)}(x,t)>=const;\nonumber
\end{eqnarray}
$T^{(\uparrow\downarrow)}$ and $R^{(\uparrow\downarrow)}$ are the (real)
transmission and reflection coefficients, respectively, for particles with spin up
$(\uparrow)$ and down $(\downarrow)$. Let further
$T=(T^{(\uparrow)}+T^{(\downarrow)})/2$ and $R=(R^{(\uparrow)}+R^{(\downarrow)})/2$
be quantities to describe all particles.

\subsubsection{Time evolution of the spin polarization of particles} \label{a332}

To study the time evolution of the average particle's spin, we have to find the
expectation values of the spin projections $\hat{S}_x$, $\hat{S}_y$ and $\hat{S}_z$.
Note, for any $t$
\begin{eqnarray*}
<\hat{S}_x>_{full}\equiv \frac{\hbar}{2}\sin(\theta_{full})\cos(\phi_{full})=\hbar
\cdot \Re(<\psi_{full}^{(\uparrow)}|\psi_{full}^{(\downarrow)}>);
\end{eqnarray*}
\begin{eqnarray} \label{9006}
<\hat{S}_y>_{full}\equiv \frac{\hbar}{2}\sin(\theta_{full})\sin(\phi_{full})=
\hbar\cdot \Im(<\psi_{full}^{(\uparrow)}|\psi_{full}^{(\downarrow)}>);
\end{eqnarray}
\begin{eqnarray*}
<\hat{S}_z>_{full}\equiv \frac{\hbar}{2}\cos(\theta_{full})=\frac{\hbar}{2}
\left[<\psi_{full}^{(\uparrow)}|\psi_{full}^{(\uparrow)}>
-<\psi_{full}^{(\downarrow)}|\psi_{full}^{(\downarrow)}>\right].
\end{eqnarray*}
Similar expressions are valid for transmission and reflection:
\begin{eqnarray*}
<\hat{S}_x>_{tr}=\frac{\hbar}{T}
\Re(<\psi_{tr}^{(\uparrow)}|\psi_{tr}^{(\downarrow)}>),\ooo
<\hat{S}_y>_{tr}=\frac{\hbar}{T}
\Im(<\psi_{tr}^{(\uparrow)}|\psi_{tr}^{(\downarrow)}>),\\
<\hat{S}_z>_{tr}=\frac{\hbar}{2T}
\Big(<\psi_{tr}^{(\uparrow)}|\psi_{tr}^{(\uparrow)}>
-<\psi_{tr}^{(\downarrow)}|\psi_{tr}^{(\downarrow)}>\Big),
\end{eqnarray*}
\begin{eqnarray*}
<\hat{S}_x>_{ref}=\frac{\hbar}{R}
\Re(<\psi_{ref}^{(\uparrow)}|\psi_{ref}^{(\downarrow)}>),\ooo
<\hat{S}_y>_{ref}=\frac{\hbar}{R}
\Im(<\psi_{ref}^{(\uparrow)}|\psi_{ref}^{(\downarrow)}>),\\
<\hat{S}_z>_{ref}=\frac{\hbar}{2R}
\left(<\psi_{ref}^{(\uparrow)}|\psi_{ref}^{(\uparrow)}>
-<\psi_{ref}^{(\downarrow)}|\psi_{ref}^{(\downarrow)}>\right).
\end{eqnarray*}

Note, $\theta_{full}=\pi/2,$ $\phi_{full}=0$ at $t=0.$ However, this is not the case
for transmission and reflection. Namely, for $t=0$ we have
\begin{eqnarray*}
\phi_{tr,ref}^{(0)}=\arctan\left(\frac{\Im(<\psi_{tr,ref}^{(\uparrow)}(x,0)|
\psi_{tr,ref}^{(\downarrow)}(x,0)>)}
{\Re(<\psi_{tr,ref}^{(\uparrow)}(x,0)|\psi_{tr,ref}^{(\downarrow)}(x,0)>)}\right);
\end{eqnarray*}
\begin{eqnarray*}
\theta_{tr,ref}^{(0)}=\arccos\Big(<\psi_{tr,ref}^{(\uparrow)}(x,0)|
\psi_{tr,ref}^{(\uparrow)}(x,0)>\\
-<\psi_{tr,ref}^{(\downarrow)}(x,0)|\psi_{tr,ref}^{(\downarrow)}(x,0)>\Big);
\end{eqnarray*}

Since the norms of $\psi_{tr}^{(\uparrow\downarrow)}(x,t)$ and
$\psi_{ref}^{(\uparrow\downarrow)}(x,t)$ are constant,
$\theta_{tr}(t)=\theta_{tr}^{(0)}$ and $\theta_{ref}(t)=\theta_{ref}^{(0)}$ for any
value of $t$. For the $z$-components of spin we have
\begin{eqnarray} \label{90018}
<\hat{S}_z>_{tr}(t)=\hbar\frac{T^{(\uparrow)}-
T^{(\downarrow)}}{T^{(\uparrow)}+T^{(\downarrow)}},\ooo
<\hat{S}_z>_{ref}(t)=\hbar\frac{R^{(\uparrow)}-
R^{(\downarrow)}}{R^{(\uparrow)}+R^{(\downarrow)}}.
\end{eqnarray}

So, since the operator $\hat{S}_z$ commutes with Hamiltonian (\ref{900200}), this
projection of the particle's spin should be constant, on the average, both for
transmission and reflection. From the most beginning the subensembles of transmitted
and reflected particles possess a nonzero average $z$-component of spin (though it
equals zero for the whole ensemble of particles, for the case considered) to be
conserved in the course of scattering. By our approach the angles
$\theta_{tr}^{(0)}$ and $\theta_{ref}^{(0)}$ cannot be used as a measure of the time
spent by a particle in the barrier region.

\subsubsection{Larmor precession caused by the infinitesimal magnetic field confined to
the barrier region} \label{a333}

As in \cite{But,Lia}, we will suppose further that the applied magnetic field is
infinitesimal. In order to introduce characteristic times let us find the
derivations $d\phi_{tr}/dt$ and $d\phi_{ref}/dt.$ For this purpose we will use the
Ehrenfest equations for the average spin of particles:
\begin{eqnarray*}
\frac{d<\hat{S}_x>_{tr}}{dt}=-\frac{\hbar\omega_L}{T} \int_a^b
\Im[(\psi_{tr}^{(\uparrow)}(x,t))^*\psi_{tr}^{(\downarrow)}(x,t)]dx\\
\frac{d<\hat{S}_y>_{tr}}{dt}=\frac{\hbar\omega_L}{T} \int_a^b
\Re[(\psi_{tr}^{(\uparrow)}(x,t))^*\psi_{tr}^{(\downarrow)}(x,t)]dx\\
\frac{d<\hat{S}_x>_{ref}}{dt}=-\frac{\hbar\omega_L}{R} \int_a^{x_c}
\Im[(\psi_{ref}^{(\uparrow)}(x,t))^*\psi_{ref}^{(\downarrow)}(x,t)]dx\\
\frac{d<\hat{S}_y>_{ref}}{dt}=\frac{\hbar\omega_L}{R} \int_a^{x_c}
\Re[(\psi_{ref}^{(\uparrow)}(x,t))^*\psi_{ref}^{(\downarrow)}(x,t)]dx.
\end{eqnarray*}
Note, $\phi_{tr,ref}=
\arctan\left(<\hat{S}_y>_{tr,ref}/<\hat{S}_x>_{tr,ref}\right).$ Hence, considering
that the magnetic field is infinitesimal and
$|<\hat{S}_y>_{tr,ref}|\ll|<\hat{S}_x>_{tr,ref}|,$ we have
\begin{eqnarray*}
\frac{d \phi_{tr}}{dt}=\frac{1}{<\hat{S}_x>_{tr}}\cdot
\frac{d<\hat{S}_y>_{tr}}{dt};\ooo \frac{d
\phi_{ref}}{dt}=\frac{1}{<\hat{S}_x>_{ref}}\cdot \frac{d<\hat{S}_y>_{ref}}{dt}.
\end{eqnarray*}
Then, considering the above expressions for the spin projections and their
derivatives on $t$, we obtain
\[\frac{d \phi_{tr}}{dt}=\omega_L \frac{\int_a^b
\Re[(\psi_{tr}^{(\uparrow)}(x,t))^*\psi_{tr}^{(\downarrow)}(x,t)]dx}
{\int_{-\infty}^\infty
\Re[(\psi_{tr}^{(\uparrow)}(x,t))^*\psi_{tr}^{(\downarrow)}(x,t)]dx};\] \[\frac{d
\phi_{ref}}{dt}=\omega_L \frac{\int_a^{x_c}
\Re[(\psi_{ref}^{(\uparrow)}(x,t))^*\psi_{ref}^{(\downarrow)}(x,t)]dx}
{\int_{-\infty}^{x_c}
\Re[(\psi_{ref}^{(\uparrow)}(x,t))^*\psi_{ref}^{(\downarrow)}(x,t)]dx}.\] Or, taking
into account that in the first order approximation on $\omega_L$, when
$\psi_{tr}^{(\uparrow)}(x,t)=\psi_{tr}^{(\downarrow)}(x,t)=\psi_{tr}(x,t)$ and
$\psi_{ref}^{(\uparrow)}(x,t)= \psi_{ref}^{(\downarrow)}(x,t)=\psi_{ref}(x,t),$ we
have
\[\frac{d \phi_{tr}}{dt}\approx\frac{\omega_L}{{\bf T}} \int_a^b
|\psi_{tr}(x,t)|^2dx;\ooo \frac{d \phi_{ref}}{dt}\approx\frac{\omega_L}{{\bf R}}
\int_a^{x_c} |\psi_{ref}(x,t)|^2dx;\] note, in this limit, $T\to{\bf T}$ and
$R\to{\bf R}$.

As is supposed in our setting the problem, both at the initial and final instants of
time, a particle does not interact with the potential barrier and magnetic field. In
this case, without loss of exactness, the angles of rotation ($\Delta\phi_{tr}$ and
$\Delta\phi_{ref}$) of spin under the magnetic field, in the course of a completed
scattering, can be written in the form,
\begin{eqnarray} \label{90020}
\Delta\phi_{tr}=\frac{\omega_L}{{\bf T}} \int_{-\infty}^\infty dt \int_a^b
dx|\psi_{tr}(x,t)|^2,\ooa \Delta\phi_{ref}=\frac{\omega_L}{{\bf R}}
\int_{-\infty}^\infty dt \int_a^{x_c} dx|\psi_{ref}(x,t)|^2
\end{eqnarray}
On the other hand, both the quantities can be written in the form:
$\Delta\phi_{tr}=\omega_L \tau^L_{tr}$ and $\Delta\phi_{eef}=\omega_L \tau^L_{ref},$
where $\tau^L_{tr}$ and $\tau^L_{ref}$ are the Larmor times for transmission and
reflection. Comparing these expressions with (\ref{90020}), we eventually obtain
\begin{eqnarray} \label{922}
\tau^L_{tr}=\frac{1}{{\bf T}} \int_{-\infty}^\infty dt \int_a^b
dx|\psi_{tr}(x,t)|^2,\ooo \tau^L_{ref}=\frac{1}{{\bf R}} \int_{-\infty}^\infty dt
\int_a^{x_c} dx|\psi_{ref}(x,t)|^2.
\end{eqnarray}
These are just the searched-for definitions of the Larmor times for transmission and
reflection.

Let us write down the wave packets for transmission and reflection in the form,
\begin{eqnarray*}
\psi_{tr,ref}(x,t)=\frac{1}{\sqrt{2\pi}}\int_{-\infty}^{\infty}
A_{in}(k)\psi_{tr,ref}(x,k)e^{-i E(k)t/\hbar}dk;
\end{eqnarray*}
expressions for $\psi_{tr}(x,k)$ and $\psi_{ref}(x,k)$ see in Section \ref{a2}. Then
Exps. (\ref{922}) can be rewritten in terms of dwell times (\ref{4005}) and
(\ref{40014}):
\begin{eqnarray} \label{823}
\tau^L_{tr}=\frac{1}{{\bf T}}\int_{0}^{\infty}\varpi(k) T(k)\tau^{tr}_{dwell}(k)
dk,\ooo \tau^L_{ref}=\frac{1}{{\bf R}}\int_{0}^{\infty}\varpi(k)
R(k)\tau^{ref}_{dwell}(k) dk.
\end{eqnarray}
where $\varpi(k)=|A_{in}(k)|^2-|A_{in}(-k)|^2.$ Thus, the Larmor times for
transmission and reflection are, in fact, the average values of dwell times
(\ref{4005}) and (\ref{40014}), respectively.

In the end of this section it is useful to address the rectangular barrier. For the
stationary case, in addition to Larmor times (\ref{4007}), (\ref{4009}),
(\ref{40030}) and (\ref{40031})), we present explicit expressions for the initial
angles $\theta_{tr}^{(0)}$ and $\phi_{tr}^{(0)}$. To the first order in $\omega_L$,
we have $\theta_{tr}^{(0)}=\frac{\pi}{2}-\omega_L \tau_z,$ $\phi_{tr}^{(0)}=\omega_L
\tau_0,$ $\theta_{ref}^{(0)}=\frac{\pi}{2}+\omega_L \tau_z$ and
$\phi_{tr}^{(0)}=-\omega_L \tau_0,$ where
\begin{eqnarray*}
\tau_z=\frac{m\kappa_0^2}{\hbar\kappa^2}\cdot\frac{(\kappa^2-k^2)\sinh(\kappa
d)+\kappa^2_0\kappa d\cosh(\kappa d)}{4k^2\kappa^2+\kappa_0^4\sinh^2(\kappa d)}
\sinh(\kappa d)\\
\tau_z=\frac{m\kappa_0^2}{\hbar\kappa^2}\cdot\frac{\kappa^2_0\kappa d\cos(\kappa
d)-\beta(\kappa^2+k^2)\sin(\kappa d)}{4k^2\kappa^2+\kappa_0^4\sin^2(\kappa d)}
\sin(\kappa d),
\end{eqnarray*}
for $E<V_0$ and $E\geq V_0$, respectively;
\begin{eqnarray} \label{90028}
\tau_0=\frac{2mk}{\hbar\kappa}\cdot\frac{(\kappa^2-k^2)\sinh(\kappa
d)+\kappa^2_0\kappa d\cosh(\kappa d)}{4k^2\kappa^2+\kappa_0^4\sinh^2(\kappa d)},\nonumber\\
\tau_0=\frac{2mk}{\hbar\kappa}\cdot\frac{\beta\kappa^2_0\kappa d\cos(\kappa
d)-(\kappa^2+k^2)\sin(\kappa d)}{4k^2\kappa^2+\kappa_0^4\sin^2(\kappa d)},
\end{eqnarray}
for $E<V_0$ and $E\geq V_0$, respectively.

Note that $\tau_z$ is just the characteristic time introduced in \cite{But} (see
Exp. (2.20a)). However, we have to stress once more that this quantity does not
describe the duration of the scattering process (see the end of Section \ref{a332}).
As regards $\tau_0,$ this quantity is directly associated with timing a particle in
the barrier region. It describes the {\it initial} position of the "clock-pointers"
which they have before entering a particle into this region.

\subsection{Tunneling a particle through an opaque rectangular barrier}\label{a4}

Note, the problem of scattering a particle, with a well defined energy, on an opaque
rectangular potential barrier is the most suitable case for verifying our approach.
Let us denote final measured azimuthal angles, for transmission and reflection, as
$\phi_{tr}^{(\infty)}$ and $\phi_{ref}^{(\infty)},$ respectively. By our approach
$\phi_{tr,ref}^{(\infty)}=\phi_{tr,ref}^{(0)}+\Delta\phi_{tr,ref}$. That is, the
final times are expected to be registered by the Larmor clock, for transmission and
reflection, should be equal to $\tau^L_{tr}+\tau_0$ and $\tau^L_{ref}-\tau_0,$
respectively.

Note, for a particle scattering on an opaque rectangular barrier (when $\kappa d\gg
1$) we have $|\tau_0|\ll\tau^L_{ref}\ll\tau^L_{tr}$ (see Exps. (\ref{4007}),
(\ref{40030}) and (\ref{90028})). As is known, Smith's dwell time
$\tau^{Smith}_{dwell}$ (which coincides with the "phase" time) and Buttiker's dwell
time saturate in this case with increasing the barrier's width (see Exps. (3.2) and
(2.20b) in \cite{But}). Just this property of the tunneling times is interpreted as
the Hartman effect.

At the same time, our approach denies the existence of this effect: transmission
time (\ref{4007}) increases as exponent when $d\to\infty.$ Of course, reflection
time (\ref{40030}) is naturally to saturate in this case.

As regards the Bohmian approach, it formally denies this effect, too. It predicts
that the average time, $\tau_{dwell}^{Bohm},$ spent by a transmitted particle in the
opaque rectangular barrier is
\begin{eqnarray*}
\tau^{Bohm}_{dwell}\equiv\frac{1}{T}\tau^{Smith}_{dwell}=\frac{m}{2\hbar
k^3\kappa^3}\Big[\left(\kappa^2-k^2\right)k^2\kappa d +\kappa_0^4 \sinh(2\kappa
d)/2\Big].
\end{eqnarray*}
Thus, for $\kappa d\gg 1$ we have $\tau^{Bohm}_{dwell}/\tau^{tr}_{dwell}\sim
\cosh(\kappa d),$ i.e., \[\tau^{Bohm}_{dwell}\gg\tau^{tr}_{dwell}\gg
\tau^{Smith}_{dwell} \sim \tau^{Butt}_{dwell}.\]

As is seen, in comparison with our definition, $\tau^{Bohm}_{dwell}$ overestimates
the duration of dwelling transmitted particles in the barrier region. Of course, at
this point we can remind that the existing Bohmian model of the scattering process
is inconsistent, since it contains nonlocality.

However, it is useful also to point out that $\tau_{dwell}^{Bohm}$ to describe
transmission was obtained in terms of $\psi_{full}.$ One can show that the input of
to-be-reflected particles into $\int_a^b|\psi_{full}(x,k)|^2 dx$ is dominant inside
the region of an opaque potential barrier. Therefore treating this time scale as the
characteristic time for transmission has no basis.

So, we state that the "causal" trajectories of transmitted and reflected particles
introduced in the Bohmian mechanics are, in fact, ill-defined. However, we have to
stress that our approach does not at all deny the Bohmian mechanics. It rather says
that "causal" trajectories for scattered particles should be redefined. Indeed, an
incident particle should have two possibility (both to be transmitted and to
reflected by the barrier) irrespective of the location of its starting point. This
means that just two causal trajectories should evolve from each staring point. Both
sets of causal trajectories must be defined on the basis of $\psi_{tr}(x,t)$ and
$\psi_{ref}(x,t).$ As to the rest, all mathematical tools developed in the Bohmian
mechanics (see, e.g., \cite{Bo1,Kr1}) remain in force.

In the end of this section it is very important to stress that the group
transmission and reflection times (see \cite{Ch5}), which are coincident for
symmetric potential barriers, lead to the Hartman effect, as the previous
approaches. Thus, our model reveals a deep difference between the dwell and group
times. Only one of them has a physical sense, and the Larmor-clock timing procedure
resolves this dilemma in favor of the former. As regards the group time, it cannot
be measured for scattering particles. And, hence, it says nothing about the
effective velocity of passing a particle (signal, information) through the barrier
region.

\section{Conclusion}

It is shown that a 1D completed scattering can be considered as an entanglement of
two alternative sub-processes, transmission and reflection, macroscopically distinct
at the final stage of scattering. For this quantum process, the (entangled) state of
the whole ensemble of particles can be uniquely presented as a sum of two solutions
to the Schr\"odinger equation to possess all needed information about the time
evolution of either sub-process, at all stages of scattering.

We develop the Larmor timing procedure to allow measuring the average time spent by
particles, of either subensemble, in the barrier region. This procedure shows that
namely the dwell time gives the time spent, on the average, by a particle in the
barrier region. As regards the group time, for scattering particles it cannot be
measured with the Larmor clock, and, hence, it has no physical sense in this case.

\end{document}